\documentclass{aa}
\usepackage{graphicx}
\begin{document}
\title{Discrete Components in the Radial Velocities of 83 ScI Galaxies: Examining the Accuracy of Tully-Fisher Distances}
\author{M.B. Bell\inst{1}, S.P. Comeau\inst{1} \and D.G. Russell\inst{2}}
\institute{Herzberg Institute of Astrophysics,
National Research Council of Canada, 100 Sussex Drive, Ottawa,
ON, Canada K1A 0R6
	\and
Owego Free Academy, Owego, NY 13827, USA}

\date{Received /Accepted }
\abstract{

Our previous analyses of the redshifts and distances of galaxies included 55 spiral galaxies and 36 Type Ia Supernovae galaxies in three separate groups. All groups showed evidence that their radial velocities contain the same discrete velocity components that are predicted from the early work of Tifft. Here we examine another new source sample containing 83 ScI galaxies. When B-band distances are used, we find clear evidence that the discrete velocities defined by Tifft's most common velocity period are present in the radial velocities of these galaxies. Assuming this is true, we are then able to use the visibility, or ease of detection, of the discrete components as an indication of the distance accuracy of the galaxies, comparing B- and I-band distances, galaxy inclination ranges, and galaxy rotation velocity ranges. We find (a), that, contrary to expectation, B-band distances for these sources are considerably more accurate than I-band, (b), that the distance accuracy varies with galaxy inclination, as expected, with the most consistent distances obtained for inclinations between 48 and 65 degrees, and (c), that distances are probably more accurate for small rotation velocities. With these results for the 83 ScI galaxies, discrete velocity components have now been detected in all 174 galaxies examined.

}
\titlerunning{Discrete velocities in galaxies} 
\authorrunning{Bell et al.}
\maketitle

\section{Introduction.}

Tifft (\cite{tif96,tif97}) has found what appear to be "discrete velocity periods" present in galaxy clusters. The most prevalent of these was found in what he referred to as common spiral galaxies and showed discrete velocity components near 36, 72, 145, 290, etc., km s$^{-1}$ (hereafter Tifft's T=0 period). Other period groups were also seen. Although less common, they all showed this same octave-related, or doubling nature. We have previously examined the radial velocities and distances of 55 Spiral galaxies and 36 Type Ia Supernovae (SneIa) galaxies with accurate distances (Bell \& Comeau \cite{bel03b,bel03c}). Both groups showed evidence for discrete, intrinsic velocity components related to those reported by Tifft. The entire radial velocity was not quantized, as thought by Tifft to be the case. Instead, it now became apparent that the discrete components \em were superimposed on top of the Hubble flow. \em They were always positive, as expected, since there is no known mechanism for producing non-Doppler blueshifts. When they were removed from the measured radial velocities, a Hubble constant of H$_{\rm o}$ = 58 km s$^{-1}$ Mpc$^{-1}$ was obtained. This value is 20\% lower than that obtained in the Hubble Key Project (Freedman et al. \cite{fre01}), and the velocity-distance plot contained less scatter. The difference between the two Hubble constants derived is due to the fact that these investigators interpreted the scatter on the Hubble plot as random peculiar velocities containing both positive and negative components.

Here we examine the radial velocities of a new group of sources containing 83 ScI galaxies. B- and I-band Tully-Fisher (TF) distances, inclinations, and rotation velocities for these galaxies are listed in Table VI of Russell (\cite{rus03}). We propose here that Tifft's discrete velocity components can be detected in these galaxies, and that we can use their visibility, or detectability, as a measure of the accuracy of the distances. Using this approach, because of the large number of sources, we can study, (a), the relative accuracy of the B- and I-band distances, (b), the accuracy of the inclination corrections as a function of inclination angle and, c), the accuracy of the rotational velocity corrections. 

\section{Analysis}

As in previous analyses (Bell \& Comeau \cite{bel03b,bel03c}), the radial velocity (V$_{\rm CMB}$) of each source was first plotted vs distance. To demonstrate this, Fig 1 shows such a plot using distances and velocities that were artificially generated with zero distance errors, no peculiar velocities, and a Hubble slope of H$_{\rm o}$ = 58. The V$_{\rm CMB}$ velocities here contain only the Hubble component plus several of Tifft's T=0 discrete velocity components superimposed. The latter are shown by the solid lines with offsets of 145, 290, 580, 1158, 2314, and 4610 km s$^{-1}$. The purpose of this data set is simply to demonstrate what the reults of our analysis will look like when Tifft's T=0 velocity components are clearly present in the data. 

The RMS deviation of the V$_{\rm CMB}$ velocities from their nearest discrete velocity lines in Fig 1 was then calculated for a range of Hubble slopes. As the Hubble slope is varied, if discrete components are present in the data, a dip in the RMS deviation values will occur at the Hubble slope where the sources coincide with the discrete velocity lines. This occurs in Fig 1 for a Hubble slope of 58. If there are no discrete velocities present in the data, a broad RMS minimum will still be obtained for that H$_{\rm o}$-value that places the densest portion of the discrete velocity lines on top of the densest portion of the source distribution. With discrete velocities present in the data, as in Fig 1, if they are always positive, as assumed here by definition, the RMS dip obtained when the source velocities coincide with the discrete velocity lines must occur at an H$_{\rm o}$-value that is slightly below that obtained for the broad minimum.

The RMS deviation in V$_{\rm CMB}$ about the T=0 discrete velocity lines in Fig 1, is shown in Fig 2 plotted as a function of H$_{\rm o}$. The two RMS dips discussed above are clearly visible, one at H$_{\rm o}$ = 58 and one near 75. The dashed line in Fig 2 shows how the curve is expected to appear if there are no discrete velocities present. In this case only one RMS dip is visible.
The curve has a slightly steeper slope on the high-H$_{\rm o}$ side (H$_{\rm o} > 70$), due to the fact that there are no longer any grid lines to fit to at high H$_{\rm o}$ values. In real data there will be associated distance uncertainties and small peculiar velocities, and as these increase, the RMS dip is expected to become broader and the minimum less deep. 

Since we have previously detected discrete velocities at a Hubble slope of 58 km s$^{-1}$ Mpc$^{-1}$, discrete velocities in the 83 ScI galaxies studied here will be assumed to have been confirmed only if an RMS deviation dip is seen near this same Hubble slope. 

\begin{figure}
\centering
\includegraphics[width=7cm]{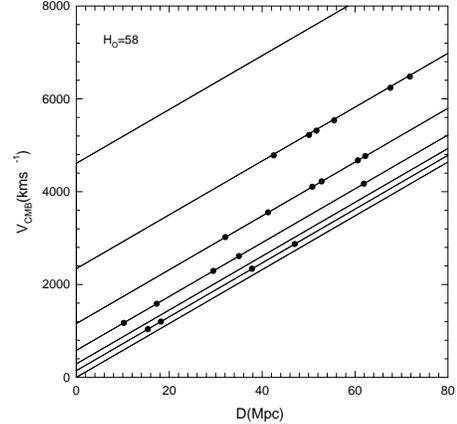}
\caption{\scriptsize{Test set of V$_{\rm CMB}$ velocities plotted vs distance. Solid lines represent Tifft's T=0 discrete velocities. \label{fig1}}}
\end{figure}

\begin{figure}
\centering
\includegraphics[width=7cm]{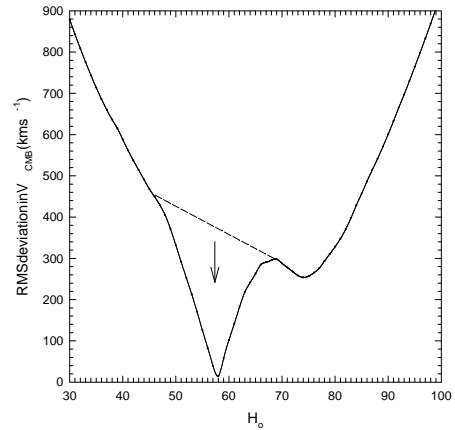}
\caption{\scriptsize{RMS deviation in test V$_{\rm CMB}$ velocities about Tifft's T=0 discrete velocity lines. \label{fig2}}}
\end{figure}

\section{Previous results for SneIa galaxies}

Fig 3 shows the result obtained previously by us (Bell \& Comeau \cite{bel03b}), using 36 SneIa galaxies (Freedman et al. \cite{fre01}). This plot shows good similarity to Fig 2, showing a clear RMS dip near H$_{\rm o}$ = 58, indicated by the arrow. Again, the dip occurs below the broad minimum produced when the densest portion of the discrete velocity distribution coincides with the densest portion of the source distribution. In this case two of Tifft's velocity period groups were included, since this appeared to be indicated by the source distribution.

\section{RMS deviation in V$_{\rm CMB}$ from the discrete velocity lines for 83 ScI galaxies}.

Fig 4 shows the result obtained using all 83 ScI galaxies and B-band distances. We assumed that only those discrete velocities present in Tifft's T=0 group were present. This velocity group was associated with common spirals, and it also corresponds to our $N$ = 1 group discussed in previous papers (Bell \cite{bel02a,bel02b,bel02c,bel02d,bel03b}). The arrow in Fig 4 indicates a dip in the RMS deviation value at a Hubble slope near H$_{\rm o}$ = 58, as found for our previous three galaxy source samples. We interpret this as an indication \em that Tifft's T=0 discrete velocities are present in the radial velocities of the 83 ScI galaxies. \em

\begin{figure}
\centering
\includegraphics[width=7cm]{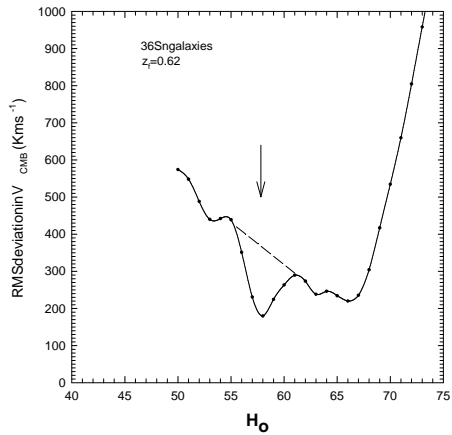}
\caption{\scriptsize{RMS deviation in V$_{\rm CMB}$ about the relevant discrete velocity lines vs H$_{\rm o}$ for SneIa galaxies. \label{fig3}}}
\end{figure}

\begin{figure}
\centering
\includegraphics[width=7cm]{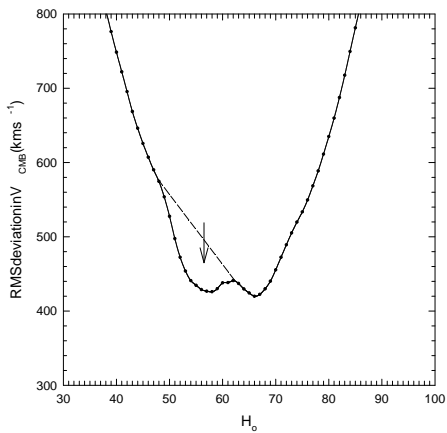}
\caption{\scriptsize{RMS deviation in V$_{\rm CMB}$ about the discrete velocity lines for all 83 ScI galaxies plotted vs H$_{\rm o}$, and using B-Band distances. \label{fig4}}}
\end{figure}

\subsection{Comparing B- and I-band distances for the ScI galaxies}

Fig 5 shows the plots obtained for the RMS deviation in V$_{\rm CMB}$ velocities for all 83 ScI galaxies for both B- and I-band TF distances. The dotted curve is for distances obtained using the I-band magnitudes. There is no evidence in this curve for an RMS deviation dip at H$_{\rm o}$ = 58, and the lowest RMS deviation found is near 580 km s$^{-1}$. The lower, solid curve is for distances obtained using the B-band magnitudes. Here, as was seen above, there is a clear dip near H$_{\rm o}$ = 58, indicated by the arrow, and its minimum RMS deviation is now close to 420 km s$^{-1}$. This fit is still poor compared to the SneIa galaxies and we interpret this as an indication that even the B-band distances of the ScI galaxies are less accurate than the SneIa distances.

\begin{figure}
\centering
\includegraphics[width=7cm]{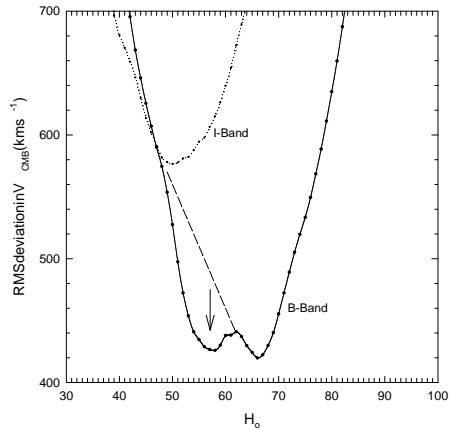}
\caption{\scriptsize{RMS deviation in V$_{\rm CMB}$ about the discrete velocity lines for all 83 ScI galaxies when (dotted curve) using I-Band distances and, (solid curve) using B-Band distances. \label{fig5}}}
\end{figure}

\begin{figure}
\centering
\includegraphics[width=7cm]{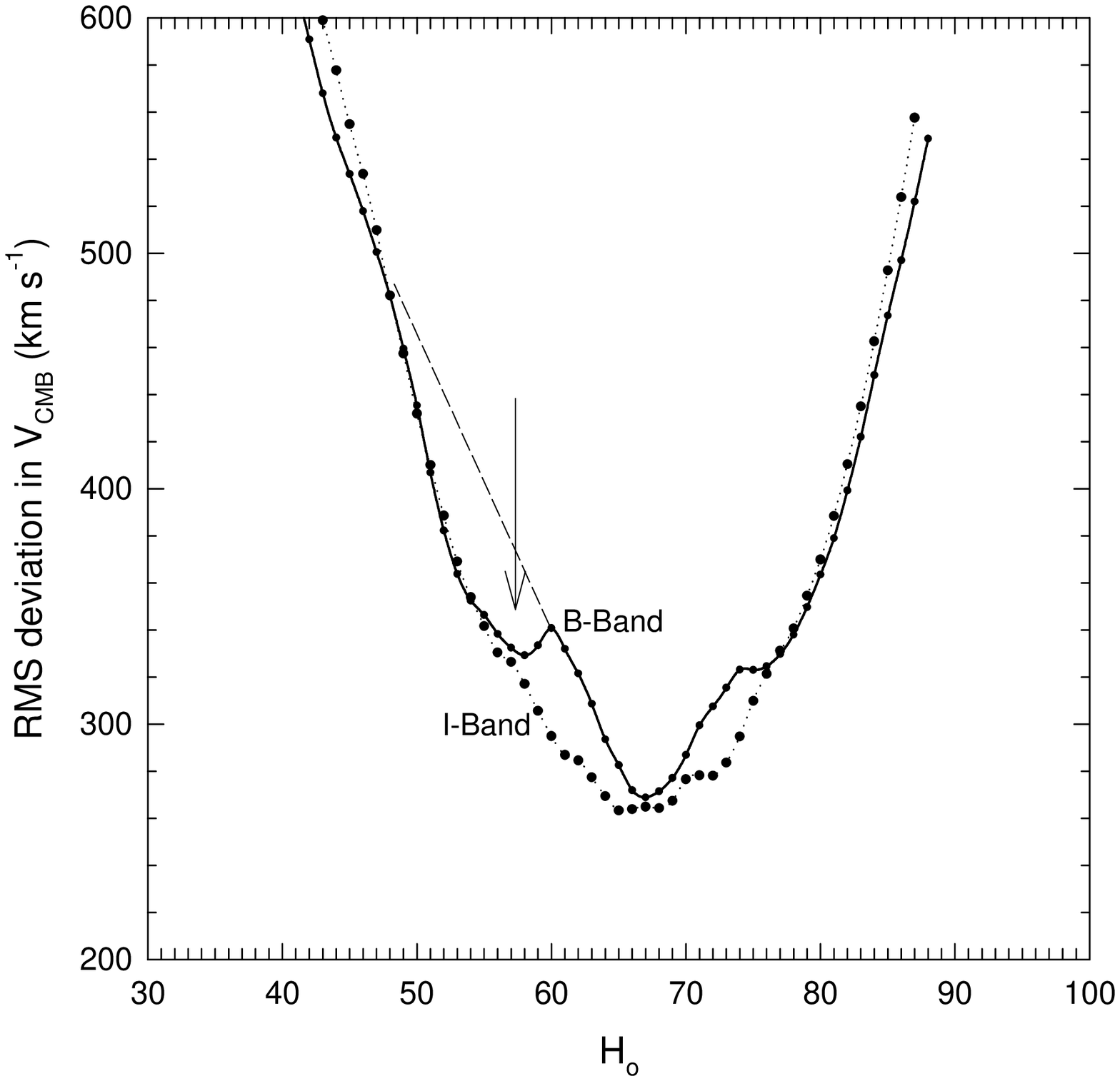}
\caption{\scriptsize{Same as Fig 5 for 60 ScI galaxies with smallest B-I distance differences, when (dotted curve) using I-Band distances and, (solid curve) using B-Band distances. \label{fig6}}}
\end{figure}

\begin{figure}
\centering
\includegraphics[width=7cm]{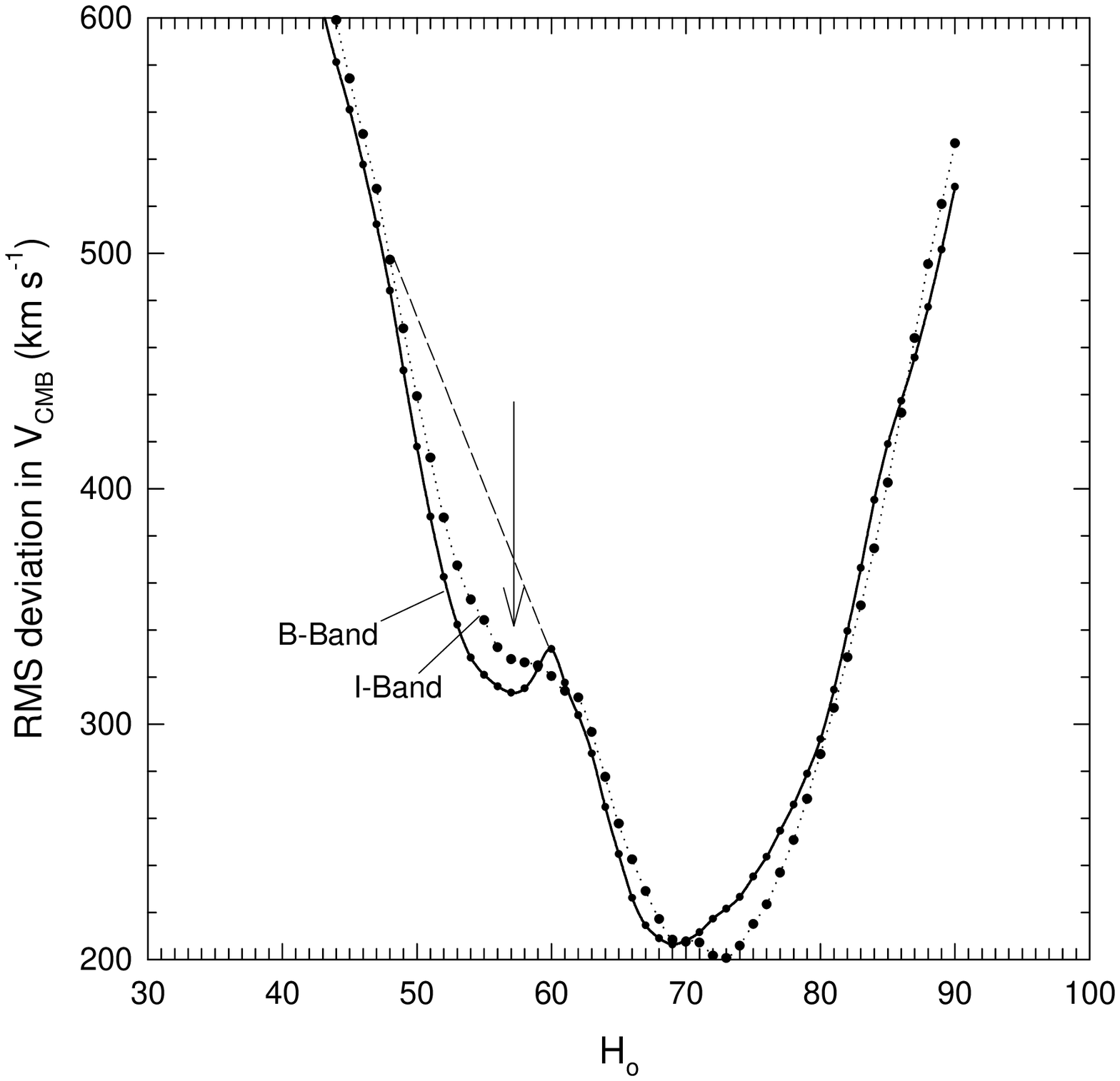}
\caption{\scriptsize{Same as Fig  for 40 ScI galaxies with smallest B-I distance differences when (dotted curve) using I-Band distances and, (solid curve) using B-Band distances. \label{fig7}}}
\end{figure}

\begin{figure}
\centering
\includegraphics[width=7cm]{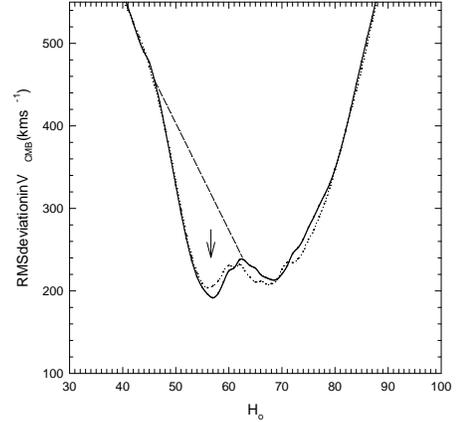}
\caption{\scriptsize{Plot of RMS deviation in V$_{\rm CMB}$ about the discrete velocity lines vs H$_{\rm o}$ using (solid)B-band and (dotted) I-band distances for 23 galaxies with smallest B-I differences. \label{fig8}}}
\end{figure}

To investigate the relative accuracy of the B- and I-band distances further, different subsets of the 83 galaxies were used. In Fig 6 the 60 galaxies with smallest B-I distance differences were used. In Fig 7 the 40 galaxies with smallest B-I distances were used, and in Fig 8 the 23 sources with smallest B-I distance differences were used. In Figs 5 through 8 there is clear evidence in each figure for an RMS deviation dip near H$_{\rm o}$ = 58 when B-band distances are used. However, only as the sources with large B-I distance differences are gradually removed from the sample does the RMS deviation dip start to appear when I-Band distances are used. Finally, Fig 8, the sample containing the 23 sources with the most similar B- and I-band distances, shows that there is essentially no difference between the two curves. \em We conclude from this that, when the B- and I-band distances differ, the B-band distances are clearly more accurate. \em 

That the B-band TF relation should provide more accurate distances than the I-band TF relation is a surprising result, at odds with standard expectations.  Bothun \& Mould (\cite{bot87}) proposed that the I-band TF relation should be more reliable, in part because the B-band is thought to be susceptible to scatter introduced from star formation events.  This result has important implications for the extra-galactic distance scale as major recent Tully-Fisher investigations have relied heavily upon I-band photometry (eg. Giovanelli et al. \cite{gio97}; Sakai et al. \cite{sak00}; Tully \& Pierce \cite{tul00}).

Unless otherwise indicated we use only B-band distances in the analysis below.

\begin{figure}
\centering
\includegraphics[width=8cm]{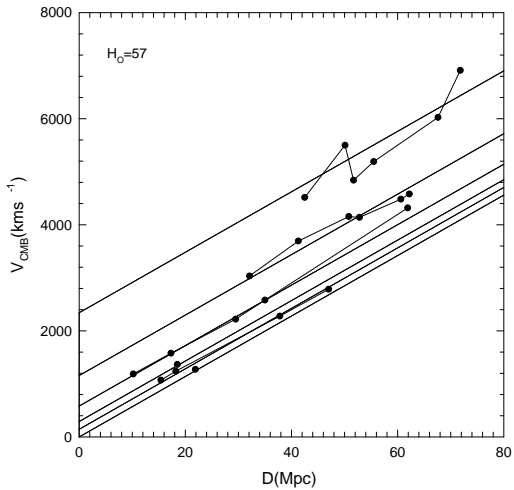}
\caption{\scriptsize{Plot of V$_{\rm CMB}$ vs B-band distance for 23 galaxies with smallest B-I distance differences in Fig 8. \label{fig9}}}
\end{figure}

\begin{figure}
\centering
\includegraphics[width=7cm]{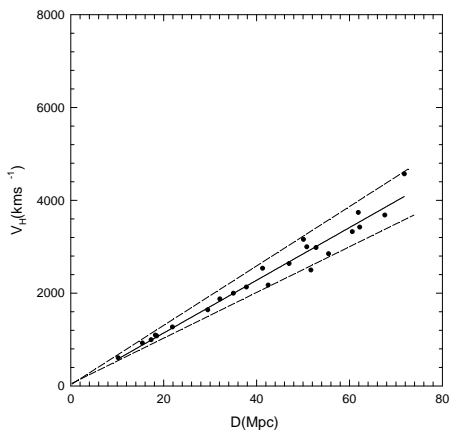}
\caption{\scriptsize{Plot of the residual velocity V$_{\rm H}$ vs B-band distance after removal of the discrete velocity components in Fig 9. \label{fig10}}}
\end{figure}

\section{V$_{\rm CMB}$ vs B-band Distance for 23 sources}

In Fig 9 the V$_{\rm CMB}$ velocities of the 23 sources in Fig 8 are plotted vs their B-band distances. The intrinsic velocity lines (Tifft's T=0 discrete velocities of 145, 290, 580, 1058, and 2314 km s$^{-1}$) are included on the plot. A solid line joins sources assumed to contain the same discrete velocity component. It is apparent from this plot that the sources are aligned along the lines representing the discrete velocity components for a Hubble slope near 57. It is also worth pointing out in Fig 9 that the mean value of the distance (D) of the sources in each discrete velocity group decreases as the discrete velocity decreases. This same characteristic has been seen in previous studies (Bell \& Comeau \cite{bel03b}), and, since we look back in time with increasing distance, it could be interpreted as a decrease in discrete velocity, or redshift, with time.

Also apparent in Fig 9 is the fact that the distance errors appear to increase as the intrinsic component increases. This can also be explained in part by the fact that the mean distance in each discrete velocity group increases as the magnitude of the discrete component increases, and the distance uncertainty is expected to increase with distance. This can be seen clearly in Fig 10 where the discrete velocity components have been removed, leaving only the component V$_{\rm H}$, due to the Hubble flow, with its uncertainty in distance and any small peculiar velocity remaining. The scatter in distance modulus for the 23 galaxies in Fig 10 is 0.14 magnitude (1 sigma) if H$_{\rm o}$ = 58, assuming no peculiar velocities. This is consistent with the cepheid calibrators for which the 12 ScI group calibrators had a B-band scatter of 0.09 magnitude (1 sigma). The 23 galaxies in Fig 10 represent those galaxies in the sample that are likely to have the smallest distance uncertainties. The scatter obtained for the entire sample of 83 galaxies will be significantly larger. Typical scatter, when type dependence is not accounted for is at least 0.25 magnitude (see Russell \cite{rus03} for a more complete discussion).

The solid line in Fig 10 represents a linear fit to the data which gave a Hubble constant of H$_{\rm o} = 57\pm2$.
Previous analyses of galaxies gave values of H$_{\rm o}$ = 57.9 for SneIa galaxies (Bell \& Comeau \cite{bel03c}), 57.5 for Sb galaxies and 60.0 for Sc galaxies (Bell \& Comeau \cite{bel03b}). With equal weighting these results give a mean value of H$_{\rm o}$ = 58.1$ \pm$1.2 km s$^{\rm -1}$ Mpc$^{\rm -1}$.

\section{Adjustments to the Magnitudes and Velocity Widths in Tully-Fisher Distance Calculations}

As pointed out on the NED database site, there are careful procedures applied to the magnitudes and velocity widths that go into the TF relation. Of these adjustments, perhaps the most important is the correction for the projection of the disk on the plane of the sky. The observed velocity width is smaller by a factor sin $(i)$, where $i$ is the galaxy inclination, than the intrinsic value. Observers correct for this by estimating $i$ from the apparent ellipticity of the galaxy disk. Formulae that make these corrections are highly susceptible to large errors when estimated inclinations are incorrect and it is well known that inclination corrections to the widths can go seriously awry at small inclinations ($i \leq40$ deg). For this reason many TF investigations use an inclination cutoff of 40-45 degrees (eg. Sakai et al. \cite{sak00}; Tully \& Pierce \cite{tul00}).

Another tricky detail of the TF relation is correcting for internal extinction. As a spiral galaxy tilts toward edge-on ($i = 90$ deg) it becomes fainter. Since spirals are viewed at a range of orientations, it is also important to correct for this effect.

Although not considered here, other corrections are also made using equations that account for morphological and luminosity class type dependence.

\begin{figure}
\centering
\includegraphics[width=7cm]{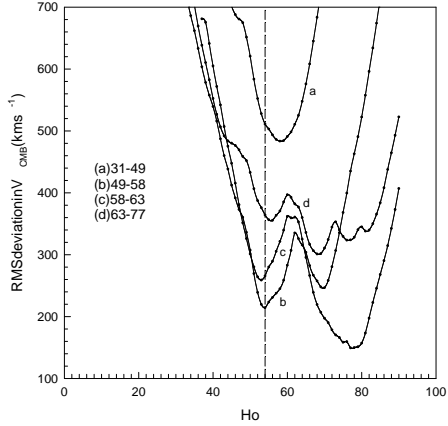}
\caption{\scriptsize{Plot of the RMS deviation in V$_{\rm CMB}$ about the discrete velocity lines vs H$_{\rm o}$ for the four inclination ranges indicated. \label{fig11}}}
\end{figure}

\begin{figure}
\centering
\includegraphics[width=7cm]{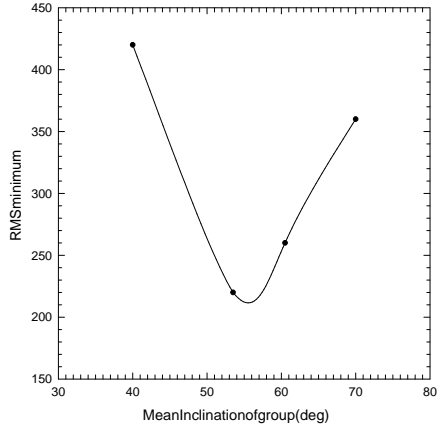}
\caption{\scriptsize{Plot of the minimum of the RMS dip near 58 vs inclination. \label{fig12}}}
\end{figure}

If the distance accuracy is a function of inclination, can we detect this?
We suggest here that it may now be possible to use the visibility of the discrete velocity components as a means of checking the relative accuracy of the magnitude corrections at various inclination angles and rotation velocities.


\begin{figure}
\centering
\includegraphics[width=7cm]{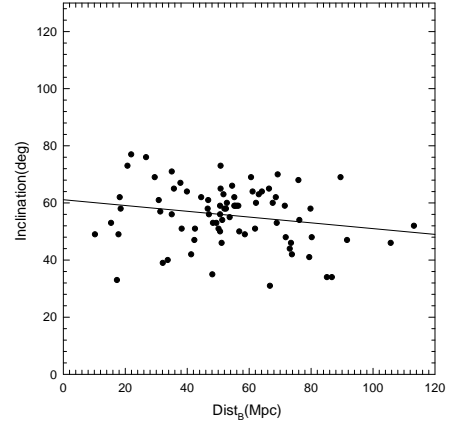}
\caption{\scriptsize{Plot of inclination vs B-band distance. \label{fig13}}}
\end{figure}

\begin{figure}
\centering
\includegraphics[width=7cm]{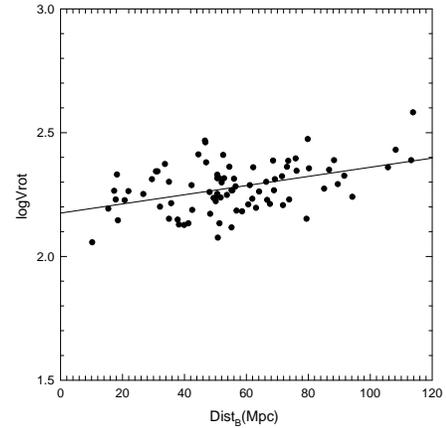}
\caption{\scriptsize{Plot of log Vrot vs B-band distance. \label{fig14}}}
\end{figure}

\begin{figure}
\centering
\includegraphics[width=7cm]{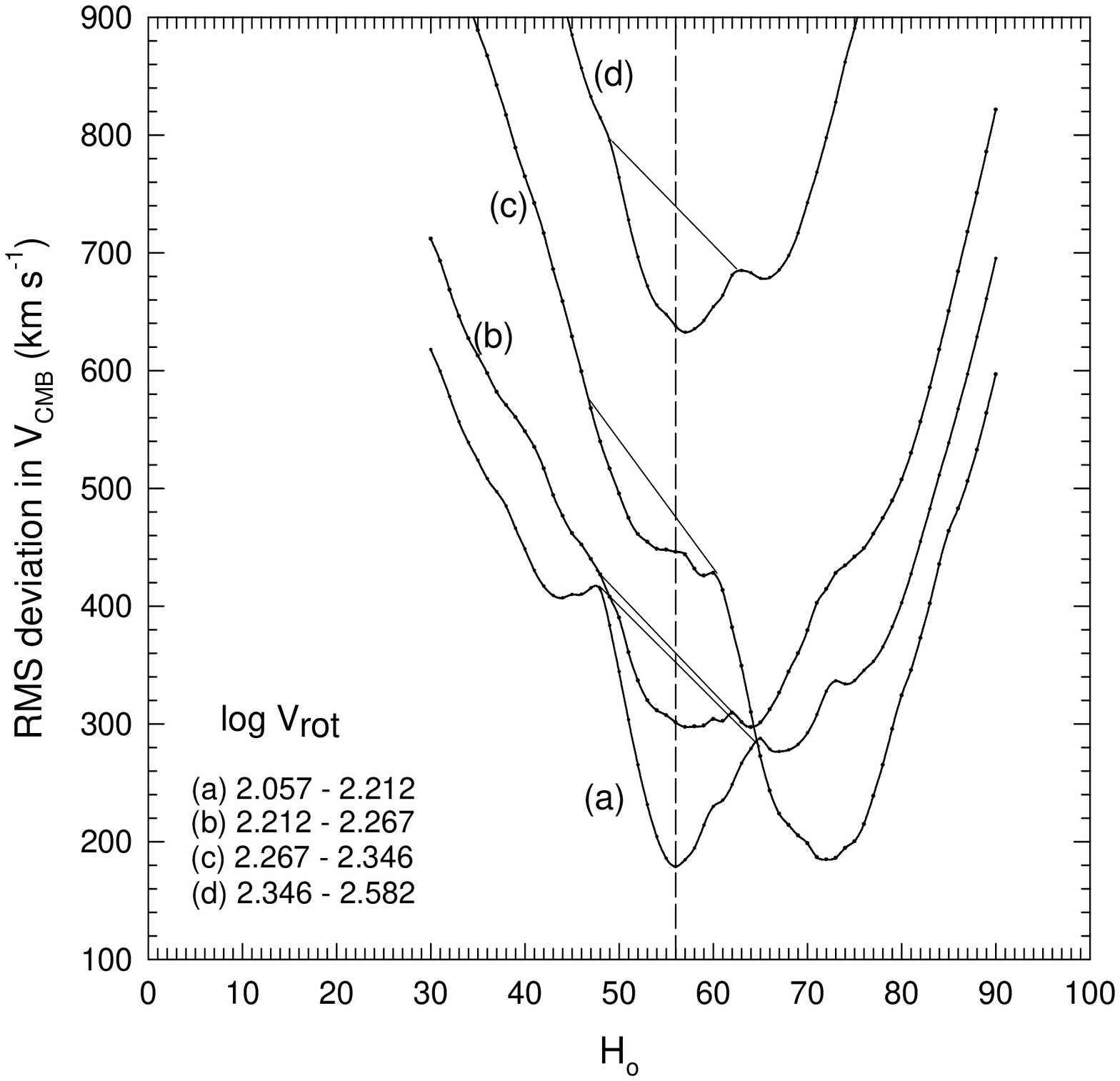}
\caption{\scriptsize{RMS deviation plots of sources in four rotation rotation velocity groups. \label{fig15}}}
\end{figure}

\section{Visibility of the Discrete Velocity Components at Various Inclination Ranges}

To investigate the possibility that the distance accuracy might vary with inclination we split the 83 galaxies into 4 inclination groups of 21 sources. Dividing the sources into equal-sized groups resulted in inclination ranges for the four groups of, (a)31-49, (b)49-58, (c)58-63, and (d)63-77 degrees. The RMS deviation in V$_{\rm CMB}$, plotted as a function of H$_{\rm o}$ for each group, is shown in Fig 11. There is clear evidence for an RMS deviation dip, indicated by the dashed line, in all inclination ranges except the lowest, (a)31-49 degrees. Also worth noting is the fact that each of the four inclination groups represents a completely different set of sources. The best fits are obtained for inclinations in the 49 to 63 degree range as shown in Fig 12 which is a plot of the minimum RMS deviation of the dip near 58 in each inclination range. This result confirms how the relative accuracy of the magnitude corrections in various inclination ranges changes, as discussed above, and also confirms that the corrections in determining the TF distance are at least the most consistent in the inclination range between 50 and 60 degrees. This may need to be kept in mind whenever discrete velocity components are sought using TF distances.

It is also of interest to note in Figs 13 and 14 that both the inclination and rotation values vary with distance. These correlations are not completely unexpected. In Fig 13, the apparent decrease in the highest inclinations with distance can be attributed to the effects of the plate cutoff as the galaxies get fainter with distance and inclination. This may also affect the rotational velocities as seen in Fig 14.

\section{Visibility of the Discrete Velocity Components with Rotation Velocity}

In Fig. 15 the 83 sources have been divided into four groups according to their rotation velocities, and the RMS deviations in V$_{\rm CMB}$ for each group plotted vs H$_{\rm o}$. Each of the groups contains an independent set of 21 sources as was done in the inclination analysis above. Although the discrete velocities are visible in all groups, the best fit is obtained for the slowest rotators.

\section{Discussion}

One of the strongest arguments favouring our interpretation that the detection of a dip in the RMS deviation in V$_{\rm CMB}$ is due to discrete velocities present in the data, is the fact that in the present analysis, as well as in all previous analyses where galaxies were examined, the dip has always occurred at the same Hubble slope. Although the occurrence of RMS dips due to random clumping in the source distribution cannot be ruled out, there are few arguments that can be made to explain their consistent appearance at the same Hubble slope. There would appear to be only two reasons why this might happen. One is if the discrete components are real, and the other is if the analysis procedure is somehow affecting the result. The latter case was examined extensively in a previous paper (Bell \& Comeau \cite{bel03a}), where it was clearly demonstrated that the reduction procedure did not introduce the RMS dip. In this examination three different approaches were used. In one of these we first carried out a linear regression on the velocity-distance data being examined, to determine its mean slope, assuming that the scatter on the plot was due to positive and negative peculiar velocities. The polarity of each peculiar velocity determined in this way was then reversed to create a new data set with all the properties of the original except for the polarity of the peculiar velocities. This data set was then processed in a manner identical to the original. Where a strong RMS dip was seen in the original data, none was seen in the data set with reversed polarity, demonstrating clearly that the reduction procedure alone does not produce an RMS dip. The other two tests were consistent with this result. We conclude that the RMS dip at H$_{\rm o}$ = 58 that we have now detected in four completely separate galaxy groups, as well as in several different sub-samples of the 83 ScI galaxies, is due to the presence of discrete velocity components in the radial velocities of these galaxies.

\section{Conclusions}

We have examined another new source sample containing 83 ScI galaxies. We find evidence that the velocity components of Tifft's most common T=0 velocity period (with velocity components of 145, 290, 580, 1158 and 2314 km s$^{-1}$) are present in the radial velocities of these galaxies. We also use the visibility, or ease of detection, of the discrete components as an indication of the accuracy of the distances. We find, for these 83 sources, that B-band distances are considerably more accurate than I-band. Using this technique we also find evidence that the distance accuracy varies with galaxy inclination, with the most consistent distances obtained for inclinations between 45 and 65 degrees. This result confirms what has generally been excepted and reinforces our belief that we have indeed detected Tifft's most common discrete velocity period in the radial velocities of these galaxies. We also find that galaxies with smaller rotational velocities in general have more accurate distances than galaxies with larger rotational velocities. Clearly the most surprising result is that the B- and I-band distance differences appear to be due to increasing uncertainty in I-band distances only.  With this sample of 83 ScI galaxies, discrete velocity components have now been detected in all 174 galaxies looked at.

The authors thank Drs. D. McDiarmid and T. Legg for helpful comments.

\end{document}